\documentclass[12pt]{iopart}

\usepackage{setstack}
\usepackage{iopams}

\newtheorem{lemma}{Lemma}
\newtheorem{theorem}{Theorem}

\newtheorem{prop}{Proposition}

\newcommand{\La}{{\Lambda}}

\newcommand{\eqand}{{\quad \mathrm{and} \quad}}

\newcommand{\del}{{\delta}}
\newcommand{\Del}{{\Delta}}
\newcommand{\diag}{{\mathrm{diag}}}
\newcommand{\be}{\begin{equation}}
\newcommand{\ee}{\end{equation}}
\newcommand{\bea}{\begin{eqnarray}}
\newcommand{\eea}{\end{eqnarray}}
\newcommand{\al}{\alpha}
\newcommand{\ba}{\beta}

\newcommand{\lb}{{\ell}}
\newcommand{\nb}{{n}}
\newcommand{\mb}[1]{{m_{(#1)}}}
\newcommand{\eb}{{e}}

\newcommand{\M}[1]{{\overset{#1}{M}}}
\newcommand{\mwand}{{multiple WAND}}

\newcommand{\Scal}{{\mathcal{S}}}
\newcommand{\hi}{\hat{i}}
\newcommand{\hj}{\hat{j}}
\newcommand{\hk}{\hat{k}}
\newcommand{\hl}{\hat{l}}

\begin{document}
\title{A higher-dimensional generalization of the geodesic part of the Goldberg-Sachs theorem}
\author{Mark Durkee and Harvey S. Reall}
\address{DAMTP, University of Cambridge\\
Centre for Mathematical Sciences,\\
Wilberforce Road, Cambridge, CB3 0WA, UK\\
M.N.Durkee@damtp.cam.ac.uk, H.S.Reall@damtp.cam.ac.uk}


\begin{abstract}
In more than four spacetime dimensions, a multiple Weyl-aligned null direction (WAND) need not be geodesic. It is proved that any higher-dimensional Einstein spacetime admitting a non-geodesic multiple WAND also admits a geodesic multiple WAND. All five-dimensional Einstein spacetimes admitting a non-geodesic multiple WAND are determined.
\end{abstract}

\section{Introduction}
The Goldberg-Sachs theorem \cite{Goldberg} states that, for any 4d spacetime that is not conformally flat and satisfies the vacuum Einstein equation
\be
\label{eqn:einstein}
  R_{\mu\nu} = \La g_{\mu\nu}, 
\ee
a null vector field is a repeated principal null direction if and only if it is tangent to a shear-free null geodesic congruence.  This result is very useful when studying algebraically special spacetimes in four dimensions, an approach which has proved to be a useful technique both for classifying known solutions, and constructing new ones.  For example, use of the Goldberg-Sachs theorem was the first step in the discovery of the Kerr solution \cite{Kerr}.

Over the past decade there has been considerable interest in General Relativity in more than four dimensions, and black holes in particular. Several techniques for solving the Einstein equation have been extended to higher dimensions but, so far, methods based on algebraic classification of the Weyl tensor have received little attention. Given the importance of these methods in 4d, it seems natural to try to generalize them to higher dimensions.

Recently, Coley, Milson, Pravda and Pravdova (CMPP) have developed a method for algebraic classification of the Weyl tensor of higher-dimensional spacetimes \cite{Classification}. Their method is based on the classification of components of the Weyl tensor by their transformation properties under local Lorentz boosts.  The higher-dimensional analogue of a principal null direction is a \emph{Weyl-aligned null direction} (WAND).  The classification is reviewed in Ref. \cite{Coley:Rev}.

The CMPP classification reduces to the standard Petrov classification in 4d, but many 4d results do not extend to higher dimensions.  For example, any 4d spacetime (that is not conformally flat) admits exactly 4 (possibly repeated) PNDs.  However, a spacetime of dimension $d>4$ might admit no WAND or it might admit infinitely many WANDs.

In this paper we shall restrict ourselves to studying Einstein spacetimes (i.e. solutions of \eref{eqn:einstein}) admitting a \emph{multiple WAND}, the higher-dimensional analogue of a repeated principal null direction. It is known that the Goldberg-Sachs theorem does not generalize to higher dimensions in an obvious way: a geodesic multiple WAND need not be shear-free \cite{Frolov}, and a multiple WAND need not be geodesic \cite{TypeD,TypeII, Mahdi}. The simplest example of the latter behaviour is a product spacetime, e.g. $dS_3 \times S^2$, where {\it any} null vector field tangent to $dS_3$ is a multiple WAND  irrespective of whether or not it is geodesic \cite{Mahdi}. However, in this example there also exist geodesic multiple WANDs. Our main result is a proof that this always happens, at least for Einstein spacetimes:
\begin{theorem}\label{thm:geo}
An Einstein spacetime admits a multiple WAND if, and only if, it admits a geodesic multiple WAND.
\end{theorem}
The ``if'' part of this theorem is trivial. To prove the ``only if'' part, we shall assume that the multiple WAND is non-geodesic and prove that there exists another multiple WAND that is geodesic. To do this, we shall prove:
\begin{theorem}\label{thm:submfd}
An Einstein spacetime that admits a non-geodesic multiple WAND is foliated by totally umbilic, constant curvature, Lorentzian, submanifolds of dimension three or greater, and any null vector field tangent to the leaves of the foliation is a multiple WAND. 
\end{theorem}
Recall that a submanifold is ``totally umbilic'' if, and only if, its extrinsic curvature is proportional to  its induced metric, i.e., $K_{\mu\nu\rho} = \xi_\mu h_{\nu\rho}$, for some $\xi_\mu$ orthogonal to the submanifold, where $h_{\mu\nu}$ is the projection onto the submanifold. This property is useful because:
\begin{lemma}\label{lem:umbilic}
A Lorentzian submanifold is totally umbilic if, and only if, it is ``totally null geodesic'', i.e., any null geodesic of the submanifold is also a geodesic of the full spacetime. 
\end{lemma}
Hence any {\it geodesic} null vector field in the constant curvature submanifolds of theorem \ref{thm:submfd} is a geodesic multiple WAND of the full spacetime, so theorem \ref{thm:geo} follows.

For the special case of five dimensions, as well as theorems \ref{thm:geo} and \ref{thm:submfd}, we have the stronger result:
\begin{theorem}\label{thm:list}
  A five-dimensional Einstein spacetime admits a non-geodesic multiple WAND if, and only if, it is locally isometric to one of the following:
  \begin{enumerate}
    \item Minkowski, de Sitter, or anti-de Sitter spacetime 
    \item A direct product $dS_3 \times S^2$ or $AdS_3 \times H^2$ 
    \item A spacetime with metric 
              \begin{equation*}
            ds^2 = r^2 d\tilde{s}_3^2 + \frac{dr^2}{U(r)} + U(r) dz^2, \qquad  U(r) = k - \frac{m}{r^2} -  \frac{\La}{4}r^2, 
                      \end{equation*}
where $m \ne 0$, $k \in \{1,0,-1\}$, $d\tilde{s}_3^2$ is the metric of a 3d Lorentzian space of constant curvature (i.e. 3d Minkowski or (anti-)de Sitter) with Ricci scalar $6k$, and the coordinate $r$ takes values such that $U(r)>0$.
 \end{enumerate}
\end{theorem}
Note that (ii) and (iii) are type D. Both admit 3d Lorentzian submanifolds of constant curvature, in agreement with theorem \ref{thm:submfd}. Solution (iii) is an analytically continued version of the 5d Schwarzschild solution\footnote{
It is a higher-dimensional generalization of the 4d B-metrics.} (generalized to allow for a cosmological constant and planar or hyperbolic symmetry). Special cases of (iii) are the Kaluza-Klein bubble of Ref. \cite{witten1} and the anti-de Sitter soliton of Ref. \cite{adssoliton}.

In more than five dimensions, there are many Einstein spacetimes that admit non-geodesic multiple WANDs. A large class of examples can be obtained as follows. Consider a 6d static axisymmetric solution (which need not admit a WAND):
\be
 ds^2 = - A(r,z)^2 dt^2 + B(r,z)^2 (dr^2 + dz^2) + C(r,z)^2 d\Omega^2,
\ee
where $d\Omega^2$ is the metric on a unit $S^3$. There are many solutions of the Einstein equation of this form, although the general solution is not known (unless it admits a multiple WAND \cite{Mahdi}). Now set $t=i\tau$ and analytically continue $d\Omega^2$ to the metric on 3d de Sitter space. This gives an Einstein metric for which any null vector field tangent to the 3d de Sitter space is a multiple WAND. This shows that there exist many 6d Einstein spacetimes admitting non-geodesic multiple WANDs. Obviously the same construction works in higher dimensions too.

This paper is organized as follows. In section \ref{sec:review} we describe our notation and summarize relevant previous work. In section \ref{sec:typeD}, we prove that an Einstein spacetime admitting a non-geodesic multiple WAND must satisfy the type D condition. This is the starting point for the proof of theorem \ref{thm:submfd} in section \ref{sec:submfd}, which also contains the proof of lemma \ref{lem:umbilic}. In section \ref{sec:list}, we prove theorem \ref{thm:list} and make some additional remarks about the 6d case. Most of our results are obtained from the Bianchi identity, whose components are written out in the Appendix.

\section{Notation and background}\label{sec:review}

\subsection{Notation}

We mainly follow the notation of Ref. \cite{TypeD}, which is similar to that of other papers in the field. For a $d$-dimensional spacetime, we use a basis $\{\lb \equiv \eb_{(0)}, \nb \equiv \eb_{(1)}, \mb{i}\equiv\eb_{(i)}\}$ where indices $i,j,k,\ldots$ run from $2$ to $d-1$, $\ell$ and $n$ are null vectors, $m_{(i)}$ are spacelike vectors, and the only non-vanishing scalar products are $\ell \cdot n =1$ and $m_{(i)} \cdot m_{(j)} = \delta_{ij}$. $d$-dimensional tangent space indices will be denoted $a,b,\ldots$, taking values from $0$ to $d-1$, while $\mu,\nu,\ldots$ are $d$-dimensional coordinate indices.

We shall assume that $\ell$ is a multiple WAND, which is equivalent to the statement
\be
 C_{0i0j} = C_{0ijk} = 0.
\ee
The following boost-weight zero quantities will play an important role in our analysis:
\be
 \Phi_{ij} \equiv C_{0i1j}, \qquad \Phi \equiv \Phi_{ii}.
\ee
Many other boost-weight zero components are related to them by various identities:
\begin{equation}
  \fl C_{01ij} = 2 C_{0[i|1|j]} = 2\Phi_{[ij]}, \quad
  C_{0(i|1|j)} = \Phi_{(ij)} = -\frac{1}{2} C_{ikjk},\quad
  C_{0101} = -\frac{1}{2} C_{ijij} = \Phi_{ii} \equiv \Phi.
\end{equation}
The boost-weight $-1$ Weyl components are determined by\footnote{Note that this differs from some other definitions of $\Psi$ in the literature, e.g.\ \cite{Bianchi}, by numerical factors and ordering of indices.} 
\be
\Psi_{ijk} \equiv C_{1ijk}
\ee
Note that contracting on the first and third indices gives
\begin{equation}
  \Psi_j \equiv C_{101j} = C_{1kjk}=\Psi_{kjk}. \label{eqn:psitrace}
\end{equation}
Finally, the boost-weight $-2$ Weyl components are described by the symmetric matrix
\be
  \Psi_{ij} \equiv C_{1i1j}.
\ee  
Define
\be
 L_{\mu\nu} = \nabla_\nu \ell_\mu, \qquad N_{\mu\nu} = \nabla_\nu n_\mu, \qquad \M{i}_{\mu\nu} = \nabla_\nu m_{(i)\mu},
\ee
and note that
\begin{equation}\label{eqn:ident1}
  \fl\quad\quad N_{0a} + L_{1a} = 0, \quad \M{i}_{0a} + L_{ia} = 0,
  \quad \M{i}_{1a} + N_{ia} = 0, \quad \M{i}_{ja} + \M{j}_{ia} = 0,
\end{equation}
and
\begin{equation}\label{eqn:ident2}
  L_{0a} = N_{1a} = \M{i}_{ia} = 0.
\end{equation} 
The covariant derivative along the frame vectors is given by
\begin{equation}
  D \equiv \lb \cdot \nabla, \quad \Delta \equiv \nb \cdot \nabla \eqand \del_i \equiv \mb{i} \cdot \nabla .
\end{equation}

\subsection{Background}

The vector $\ell$ is geodesic if, and only if $L_{i0} = 0$ everywhere. Throughout this paper, we shall study the case in which $\ell$ is non-geodesic:
\be
\label{eqn:Li0nonzero}
 L_{i0} \ne 0.
\ee
More precisely, we shall work in an open subset of spacetime in which equation \eref{eqn:Li0nonzero} is satisfied. We shall assume that spacetime is analytic in order to extend results from this open subset to the rest of the spacetime.\footnote{
Most work on algebraically special solutions assumes that spacetime is analytic. In smooth, but non-analytic spacetimes, the algebraic type can differ in disjoint open subsets of spacetime, even in 4d.}

In Ref. \cite{TypeD} it was proved (for Einstein spacetimes) that, if $\ell$ is a non-geodesic multiple WAND then $\Phi_{ij}$ is symmetric, and has an eigenvalue $-\Phi$ with associated eigenvector $L_{i0}$:
\be
 \Phi_{[ij]} = 0, \label{eqn:phisym}
\ee 
\be
\label{eqn:Li0evec}
 \Phi_{ij} L_{j0} = -\Phi L_{i0}.
\ee
The proofs of the results described in the introduction will be based on the Bianchi identity satisfied by the Weyl tensor (in an Einstein spacetime):
\be
 \nabla_{[e|} C_{ab|cd]} = 0.
\ee
The components of this equation in a null basis have been written out in Ref. \cite{Bianchi}. In the Appendix, we give these components, assuming that $\ell$ is a non-geodesic multiple WAND. 

\section{Satisfying the type D condition}\label{sec:typeD}

The following result will be useful in the proof of theorem \ref{thm:submfd} (note that a spacetime that satisfies the type D condition might not be type D; it might be more special):

\begin{prop}\label{prop:typeD}
An Einstein spacetime admitting a non-geodesic multiple WAND $\ell$ satisfies the type D condition with $\ell$ one of the two preferred multiple WANDs.
\end{prop}
\noindent{\it Proof:} Assume we have an Einstein spacetime with a non-geodesic \mwand\ $\ell$. We know that $L_{i0} \ne 0$. $L_{i0}$ transforms as a vector under rotations of the spatial basis vectors $m_{(i)}$. Therefore we can choose these basis vectors so that
\be
 L_{20} \ne 0, \qquad L_{\hi 0}=0,
\ee
where $\hi$, $\hj$ etc take values $3,4, \ldots , (d-1)$. From equation \eref{eqn:Li0evec}, we have
\be
 \Phi_{22} = -\Phi, \qquad \Phi_{2\hi} = 0.
\ee
Then, equation \eref{eqn:B8} gives
\be
\label{eqn:C2i2j}
 C_{2\hi 2 \hj} = \Phi_{\hi \hj} \qquad C_{2 \hi \hj \hk} = 0
\ee
and \eref{eqn:B15} with $ijkl=22 \hi \hj$ gives
\be
\label{PhiL}
 \Phi L_{[\hi \hj]} - L_{\hk [\hi} \Phi_{\hj] \hk} = 0.
\ee
Now consider \eref{eqn:B3}. The $2\hi$ component gives
\be
\label{B32i}
 \left( \Psi_{22\hi} - \Psi_{\hi} \right) L_{20} = \left( \Phi_{\hi \hj} + \Phi \delta_{\hi\hj} \right) L_{\hj 2},
\ee
and, using (\ref{PhiL}), the $\hi\hj$ component gives
\be
 \Psi_{2\hi\hj}=0.
\ee
Consider \eref{eqn:B5}. The antisymmetric part reproduces \eref{eqn:B3} so consider the symmetric part. 
Setting $i=2$, $j=\hi$ gives
\be
 \left( \Psi_{22\hi} + \Psi_{\hi} \right) L_{20} = \left( \Phi_{\hi \hj} + \Phi \delta_{\hi\hj} \right) L_{\hj 2} -2 \Phi_{\hi \hj} \stackrel{2}{M}_{\hj 0} - 2 \Phi \stackrel{2}{M}_{\hi 0}.
\ee
Subtracting this from (\ref{B32i}) gives
\be
-\Psi_{\hi} L_{20} = \Phi_{\hi \hj} \stackrel{2}{M}_{\hj 0} +  \Phi \stackrel{2}{M}_{\hi 0}.\label{eqn:psii}
\ee
Now we shall show that the basis vectors $n$, $m_{(i)}$ can be chosen to make the negative boost weight Weyl components vanish. Consider moving to a new basis $\{\lb'=\ell,\nb',\mb{i}'\}$ by performing a null rotation about $\ell$:
\be
 \ell'=\ell, \qquad n' = n - z_i \mb{i} - \frac{1}{2} z^2 \ell, \qquad \mb{i}' = \mb{i} + z_i \ell,
\ee
where $z_i$ are some smooth functions, and $z^2\equiv z_i z_i$.  In the new basis we have
\be
 \stackrel{2}{M'}_{\hi 0} = \M{2}_{\hi 0} - z_{\hi} L_{20}.
\ee
We can always choose $z_{\hi}$ so that the RHS vanishes. Hence we can always choose our basis so that (this equation is trivial for $i=2$)
\be
\label{basischoice}
 \stackrel{2}{M}_{i 0}=0. 
\ee
We shall assume this henceforth.  We now have, from \eref{eqn:psii}, that
\be
 \Psi_{\hi} = 0.
\ee
The $\hi\hj$ component of the symmetric part of \eref{eqn:B5} gives
\be
 D\Phi_{\hi \hj} = - \Phi L_{(\hi\hj)} - L_{\hk (\hi} \Phi_{\hj) \hk} - \Psi_{(\hi\hj)2} L_{20} + \stackrel{\hi}{M}_{\hk 0} \Phi_{\hk \hj} + \stackrel{\hj}{M}_{\hk 0} \Phi_{\hk \hi}.
\ee
Now consider the $2\hi 2 \hj$ component of \eref{eqn:B12}. This gives
\be
D\Phi_{\hi \hj} = - \Phi L_{(\hi\hj)} - L_{\hk (\hi} \Phi_{\hj) \hk} +2 \Psi_{(\hi\hj)2} L_{20} + \stackrel{\hi}{M}_{\hk 0} \Phi_{\hk \hj} + \stackrel{\hj}{M}_{\hk 0} \Phi_{\hk \hi}.
\ee
Comparing these two equations reveals that $\Psi_{(\hi\hj)2} =0$. However, we also have that $\Psi_{2\hi \hj}=0$, so the identity $\Psi_{[ijk]}=0$ implies that $\Psi_{[\hi\hj]2}=0$. Combining these results, we learn that
\be
 \Psi_{\hi \hj 2}=0.
\ee
Using (\ref{basischoice}), the $2\hi\hj\hk$ component of \eref{eqn:B12} reduces to $L_{20} \Psi_{\hi\hj\hk}=0$ hence
\be
 \Psi_{\hi\hj\hk}=0.
\ee
Now we have $0=\Psi_{\hi}=\Psi_{2\hi 2} + \Psi_{\hj \hi \hj} = \Psi_{2\hi 2}$. Hence all components of $\Psi_{ijk}$ vanish, and therefore so must $\Psi_i$:
\be
 \Psi_{ijk} = \Psi_i = 0.
\ee
Next, consider \eref{eqn:B6}. Setting $i=j=2$ and $k=\hi$ gives
\be
\label{Psi2i}
 \Psi_{2\hi}  L_{20} = \left( \Phi_{\hi\hj} + \Phi \delta_{\hi\hj} \right) \left( L_{\hj 1} - N_{\hj 0} \right).
 \ee
and setting $ijk=\hi\hj 2$ gives
\be
 \Psi_{\hi \hj} = 0.
\ee
The $\hi\hj\hk$ component gives
\be
 0 = \left( 2 \Phi_{\hi [ \hj} \delta _{\hk] \hl} - C_{\hi\hl\hj\hk} \right) \left( L_{\hl 1} - N_{\hl 0} \right).
\ee
Contracting on $\hi$ and $\hj$, using 
\be \label{eqn:trace}
  C_{\hi\hl\hi\hk}=C_{i\hl i\hk} - C_{2\hl 2 \hk} = -3 \Phi_{\hl\hk} \eqand \Phi_{\hi\hi} = \Phi_{ii} - \Phi_{22} = 2\Phi
\ee
gives
\be
 0 =  \left( \Phi_{\hk\hl} + \Phi \delta_{\hk\hl} \right) \left( L_{\hl 1} - N_{\hl 0} \right).
\ee
Substituting this into (\ref{Psi2i}) gives
\be
 \Psi_{2\hi}=0.
\ee
It remains to show $\Psi_{22}=0$. Setting $i=2$ in \eref{eqn:B1} gives
\be
 \delta_2 \Phi = - \Psi_{22} L_{20}.
\ee 
Setting $i=\hi$, $j=2$, $k=\hk$ in \eref{eqn:B7} and tracing on $\hi$ and $\hk$ gives
\be
 \delta_2 \Phi = - \frac{1}{2} \left( \Phi \stackrel{2}{M}_{\hi\hi} + \Phi_{\hi\hj} \stackrel{2}{M}_{\hi\hj} \right).
\ee
However, setting $m=2$ and $ijkl=\hi\hj\hk\hl$ in \eref{eqn:B16}, tracing on $\hi$ and $\hk$ and then tracing on $\hj$ and $\hl$ gives
\be
 \delta_2 \Phi = - \frac{2}{3} \left( \Phi \stackrel{2}{M}_{\hi\hi} + \Phi_{\hi\hj} \stackrel{2}{M}_{\hi\hj} \right).
\ee
Hence we conclude that
\be
 \delta_2 \Phi=0,
\ee
and hence $\Psi_{22}=0$, so
\be
 \Psi_{ij}=0.
\ee
Therefore, all of the components of the Weyl tensor of non-zero boost weight vanish, and hence the type D condition is satisfied in this basis. $\Box$

\section{Proof of theorem \ref{thm:submfd} and lemma \ref{lem:umbilic}}\label{sec:submfd}

\subsection{Proof of theorem \ref{thm:submfd}}

Assume that we have an Einstein spacetime admitting a non-geodesic multiple WAND $\ell$.
From proposition \ref{prop:typeD}, we can use a basis in which the type D condition is satisfied. Consider a new basis defined by a null rotation about $n$:
\be
\label{newbasis}
 \lb' = \lb - z_i \mb{i} - \frac{1}{2} z^2 n, \qquad n'=n, \qquad \mb{i}' = \mb{i} + z_i \lb,
\ee
where $z_i$ are arbitrary smooth functions and $z^2 \equiv z_i z_i$. Using the type D property, in the new basis we have
\be
 C'_{0ijk} = C_{iljk} z_l - 2\Phi_{i[j} z_{k]},
\ee 
\be
 C'_{0i0j} = C'_{0ijk} z_k + z_i \left( \Phi_{jk} z_k + \Phi z_j \right).
\ee
Now choose the functions $z_i$ so that $C'_{0ijk}=0$, i.e., 
\be
\label{zdef}
 C_{iljk} z_l - 2\Phi_{i[j} z_{k]}=0.
\ee
This equation certainly admits non-vanishing solutions $z_i$ because \Eref{eqn:B8} shows that $z_i = L_{i0}$ is a solution and, by our assumption \eref{eqn:Li0nonzero} that $\ell$ is non-geodesic, this solution is non-vanishing. Tracing on $i$ and $k$ reveals that $z_i$ is an eigenvector of $\Phi_{ij}$ with eigenvalue $-\Phi$:
\be
\label{Phiz}
 \Phi_{ij} z_j = -\Phi z_i.
\ee
The previous two equations imply that $C'_{0i0j}=0$. Hence for any change of basis defined by $z_i$ satisfying (\ref{zdef}), all positive boost weight Weyl components vanish  in the new basis (\ref{newbasis}), and hence $\ell'$ is a multiple WAND. Since $n'=n$ is also a multiple WAND, the negative boost weight Weyl components still vanish, and hence the type D condition is still satisfied in the new basis.

The LHS of (\ref{zdef}) defines a linear map on $z_i$ at any point in spacetime. We know that the kernel $K$ of this map is non-empty. The dimension of the kernel may vary in spacetime. Let $n\ge 1$ denote the minimum value of this dimension. The dimension of the kernel will exceed $n$ only on a set of zero measure (assuming analyticity). Let $p$ be a point at which the dimension of the kernel is $n$. By smoothness, there must be a neighbourhood of $p$ in which the dimension also equals $n$. We shall work in such a neighbourhood, and extend our results to the rest of spacetime by analyticity. In this neighbourhood, there exist $n$ linearly independent solutions $z_i$ of (\ref{zdef}), and hence a $n$-parameter family of multiple WANDs at any point. This family obviously contains $\ell$.

The $n$ solutions $z_i$ define a $n$-dimensional distribution spanned by vector fields of the form $z_i \mb{i}$. By rotating the spatial basis, we can divide it into a set $\{ \mb{I} \}$ that spans this distribution and a set $\{ \mb{\alpha} \}$ that is orthogonal to it. Here, indices $I,J,\ldots$ take values $2,3, \ldots, (n+1)$ and indices $\alpha,\beta,\ldots$ take values $(n+2),(n+3),\ldots, (d-1)$. By definition, the general solution of equation (\ref{zdef}) is
\be
 z_{\alpha} = 0,
 \ee
and $z_I$ are arbitrary functions. From equation (\ref{Phiz}), it follows that
\be
\label{eqn:PhiIJ}
 \Phi_{IJ} = -\Phi \delta_{IJ}, \qquad \Phi_{I\alpha}=0.
\ee
The vectors $\mb{\alpha}$ can be chosen to diagonalize $\Phi_{\alpha\beta}$. Note that we do {\it not} know that all eigenvalues of $\Phi_{\alpha\beta}$ differ from $-\Phi$.

In this basis, equation (\ref{zdef}) reduces to
\be
\label{eqn:Weylcpts}
 \fl C_{IJKL} = -2\Phi \delta_{I[K} \delta_{L]J}, \qquad C_{I\alpha J\beta} = \delta_{IJ} \Phi_{\alpha\beta}, \qquad 
 C_{IJK \alpha} = C_{IJ \alpha\beta} = C_{I \alpha \beta\gamma}=0.
\ee
We shall now use the Bianchi identities to deduce constraints on the form of $L_{ij}$, $N_{ij}$ and $\M{i}_{jk}$. The following will be useful:
\begin{lemma}
  If $X_\alpha$ obeys $C_{\alpha \beta\gamma \delta} X_{\delta} - 2 \Phi_{\gamma[ \alpha} X_{\beta]} =0$ everywhere then $X_\alpha = 0$ everywhere.
\end{lemma}
\noindent {\it Proof}. Extend $X_\alpha$ to $X_i$ by defining $X_I=0$. Tracing on $\alpha$ and $\gamma$ gives $\Phi_{\beta \delta} X_\delta = -\Phi X_{\delta}$. One can now check that all components of $C_{ijkl} X_l - 2\Phi_{k[i} X_{j]}$ vanish everywhere, and therefore $X_i$ lies in the kernel $K$ described above.  But the directions $\mb{\alpha}$ were defined to be those orthogonal to the kernel, and hence it follows that $X_\alpha = 0$. $\Box$

Now note that Equations \eref{eqn:B8}, and  \eref{eqn:B10} imply that
\be
\label{eqn:Lalpha0}
 L_{\alpha 0} = N_{\alpha 1} = 0.
\ee 
Equation \eref{eqn:B6} says that $L_{i1} - N_{i0}$ obeys (\ref{zdef}) everywhere hence
\be
\label{eqn:Nalpha0}
 N_{\alpha 0} = L_{\alpha 1}.
\ee
Setting $ijkl=\gamma\alpha \beta I$ in \eref{eqn:B15} gives
\be
 C_{\alpha \beta\gamma \delta} L_{\delta I} - 2 \Phi_{\gamma[ \alpha} L_{\beta]I} =0
\ee
so from the lemma (treating $I$ as fixed) we obtain
\be
 \label{eqn:LalphaI0}
 L_{\alpha I}=0.
\ee 
Similarly, from \eref{eqn:B14} we obtain
\be
\label{eqn:NalphaI0}
 N_{\alpha I} =0.
\ee
Setting $ijkl=\alpha\beta\gamma I$ in \eref{eqn:B12} and using \eref{eqn:LalphaI0} gives
\be
 C_{\alpha \beta\gamma \delta} \stackrel{I}M_{\delta 0} - 2 \Phi_{\gamma[ \alpha} \stackrel{I}{M}_{\beta]0} =0,
\ee
so the lemma gives
\be
\label{eqn:MalphaI0}
 \stackrel{I}M_{\alpha 0} =0.
\ee 
Similarly, from  \eref{eqn:B13}, we obtain 
\be
\label{eqn:MalphaI1}
 \stackrel{I}M_{\alpha 1} =0.
\ee 
Next, setting $ijklm=I\beta\delta J \gamma$ in \eref{eqn:B16} we obtain
\be
 \fl 2 \delta_{[\gamma} \Phi_{\delta] \beta} \delta_{IJ} = - 2 \Phi_{\beta [\gamma} \stackrel{I}{M}_{\delta] J} - C_{\beta \alpha \gamma\delta} \stackrel{\alpha}{M} _{IJ} + 2 \delta_{IJ} \left( \Phi_{\alpha \beta} \stackrel{\alpha}{M}_{[\gamma \delta]} + \Phi_{\alpha [\gamma} \stackrel{\alpha}{M}_{|\beta| \delta]} \right).
 \ee
However, setting $ijk=\beta\gamma\delta$ in \eref{eqn:B7} gives
 \be
   -2 \delta_{[\gamma} \Phi_{\delta] \beta} = 2 \Phi_{\beta[\gamma} L_{\delta] 1} - C_{\beta\alpha \gamma \delta} L_{\alpha 1} - 2 \left( \Phi_{\alpha \beta} \stackrel{\alpha}{M}_{[\gamma \delta]} + \Phi_{\alpha [\gamma} \stackrel{\alpha}{M}_{|\beta| \delta]} \right).
\ee 
Combining these two equations gives
\be 
 C_{\beta\alpha \gamma\delta} X_{\alpha IJ} - 2 \Phi_{\beta [\gamma} X_{\delta] IJ} = 0,
\ee 
where $X_{\alpha IJ} = L_{\alpha 1} \delta_{IJ} + \M{\alpha}_{IJ}$.  Hence, using the lemma, we have
\be\label{eqn:MalphaIJ}
 \stackrel{\alpha}{M}_{IJ} = - L_{\alpha 1} \delta_{IJ}.
\ee
A convenient way of summarizing the above results is to define indices $A,B,\ldots$ to take values $0,1,2,\ldots (n+1)$. Using equations \eref{eqn:Weylcpts} and the definition of $\Phi_{ij}$, we find that
\be
\label{eqn:submfdWeyl}
 C_{ABCD} = -2\Phi \eta_{A[C} \eta_{D]B},
\ee
where $\eta_{AB}$ is the Minkowski metric ($\eta_{01}=\eta_{10}=1$, $\eta_{IJ} = \delta_{IJ}$).
We also have (using the type D condition and equations \eref{eqn:Weylcpts})
\be
\label{eqn:submfdWeyl2}
 C_{ABC\alpha} = C_{AB\alpha\beta} = C_{A\alpha\beta\gamma}=0, \qquad C_{A\alpha B\beta}= \eta_{AB} \Phi_{\alpha\beta}.
\ee
Equations \eref{eqn:Lalpha0}, \eref{eqn:Nalpha0}, \eref{eqn:LalphaI0}, \eref{eqn:NalphaI0}, \eref{eqn:MalphaI0},  \eref{eqn:MalphaI1} and \eref{eqn:MalphaIJ} are equivalent to
\be
\label{eqn:MalphaAB}
 \M{\alpha}_{AB} = -L_{\alpha 1} \eta_{AB}.
\ee
Using this, we have
\be
\label{eqn:integrable}
 [e_{(A)},e_{(B)}]_\alpha \equiv 2 \M{\alpha}_{[AB]}=0.
\ee
Hence the distribution spanned by $\{e_{(A)} \} = \{\lb,\nb,\mb{I} \}$ is integrable, i.e., tangent to $(n+2)$-dimensional submanifolds of spacetime. From equations \eref{eqn:submfdWeyl} and \eref{eqn:submfdWeyl2}, it follows that {\it any} null vector tangent to these submanifolds is a multiple WAND.

Now the extrinsic curvature tensor of one of the submanifolds is defined by
\be
 K(X,Y) = (\nabla_X Y)^\perp,
\ee
where $X$ and $Y$ are vector fields tangent to the submanifold, and $\perp$ is the projection perpendicular to the submanifold. The non-vanishing components are
\be
 K^{\alpha}_{AB} = - \M{\alpha}_{AB} = L_{\alpha 1} \eta_{AB},
\ee
where we used \eref{eqn:MalphaAB}. Hence the submanifolds are totally umbilic.

Let $\Scal$ be one of the submanifolds. Calculating the Riemann tensor $\tilde{R}_{abcd}$ of $\Scal$ gives
\be
\label{eqn:RiemS1}
   \tilde{R}_{abcd} = h_{a}^{\phantom{a}a'} h_{b}^{\phantom{b}b'} h_{c}^{\phantom{c}c'} h_{d}^{\phantom{d}d'} \left[R_{a'b'c'd'} + 2\M{\alpha}_{c'[a'|} \M{\alpha}_{d'|b']} \right],
\ee
where
\be
  h_{ab} = \eta_{AB} e^{(A)}_a e^{(B)}_b = \eta_{ab} - m^{(\alpha)}_{a} m^{(\alpha)}_{b}
\ee
is the projection operator onto $\Scal$. Using $e_{(A)}$ as a basis on $\Scal$, we have (using the relation between the Riemann and Weyl tensors in $d$ dimensions, as well as the Einstein equation)
\be
\label{eqn:RiemS2}
\tilde{R}_{ABCD} = C_{ABCD} + \frac{2\Lambda}{d-1}  \eta_{A[C} \eta_{D]B} + 2 \M{\alpha}_{[C|A} \M{\alpha}_{|D]B}
\ee
Using equations \eref{eqn:submfdWeyl} and \eref{eqn:MalphaAB} now gives
\be
\label{eqn:projcurv}
 \tilde{R}_{ABCD} = 2 {\cal R} \eta_{A[C} \eta_{D]B},
\ee
where
\be
  {\cal R} = \frac{\Lambda}{d-1} - \Phi + L_{\alpha 1}L_{\alpha 1} .
\ee
\Eref{eqn:projcurv} is the statement that $\Scal$ has constant curvature. (The $(n+2)$-dimensional Bianchi identity implies that ${\cal R}$ is constant on $\Scal$.)  $\Box$

Note that from the $IJ$ components of equations \eref{eqn:B5}, \eref{eqn:B4} and the $IJK$ components of \eref{eqn:B7} we also have
\be
 D\Phi = \Delta\Phi = \delta_I \Phi = 0, \label{eqn:constphi}
\ee
so $\Phi$ is constant on any of the constant curvature submanifolds.

In the 7d example of Ref. \cite{TypeD}, it can be shown that the foliation is by 3d Lorentzian submanifolds of constant curvature, and that the arbitrary function appearing in this solution can be eliminated by a change of coordinates.

\subsection{Proof of lemma \ref{lem:umbilic}}

Let $\Scal$ be a Lorentzian submanifold of spacetime. Consider an affinely parametrized null geodesic of $\Scal$ with tangent vector $U$, i.e., we have $U \cdot \hat{\nabla} U= 0$, where $\hat{\nabla}$ is the Levi-Civita connection in $\Scal$. This is equivalent to $(U \cdot \nabla U)^\parallel = 0$, where $\parallel$ denotes the projection tangential to $\Scal$. Now, from the definition of the extrinsic curvature $K$, we have
\be
 (U \cdot \nabla U)^\perp = K(U,U).
\ee
If $\Scal$ is totally umbilic then the RHS vanishes because $U$ is null. Therefore all components of $U \cdot \nabla U$ vanish so $\Scal$ is totally null geodesic. Conversely, if the manifold is totally null geodesic then pick a point $p$ on $\Scal$, let $U$ be an arbitrary null vector tangent to $\Scal$ at $p$, and consider the geodesic in $\Scal$ that has tangent vector $U$ at $p$. By assumption, this is a geodesic of the full spacetime, so the RHS of the above equation must vanish. But $p$ and $U$ are arbitrary, so $K(U,U)$ must vanish for any null $U$ tangent to $\Scal$, which implies that $\Scal$ is totally umbilic.

\section{Proof of theorem \ref{thm:list} and comments on 6d case} \label{sec:list}

\subsection{Proof of theorem \ref{thm:list}}

In a 5d spacetime, the boost weight zero components of the Weyl tensor are all determined by $\Phi_{ij}$. In particular:
\begin{equation}\label{eqn:CPhi5d}
  C_{ijkl} = 2(\del_{il}\Phi_{(jk)}-\del_{ik}\Phi_{(jl)}-\del_{jl}\Phi_{(ik)}+\del_{jk}\Phi_{(il)})
                    - \Phi (\del_{il}\del_{jk} - \del_{ik} \del_{jl}).
\end{equation}
Assume that we have a non-geodesic multiple WAND $\ell$. From proposition \ref{prop:typeD}, we know that we can choose a basis $\{\ell,n,m_{(i)} \}$ so that the type D condition is satisfied. Following Ref. \cite{TypeD}, we can substitute equation \eref{eqn:CPhi5d} into \eref{eqn:B8},   to learn that the eigenvalues of $\Phi_{ij}$ must be $-\Phi,\Phi,\Phi$. Therefore we can choose the spatial basis vectors $m_{(i)}$ so that
\begin{equation}
  \Phi_{ij} = \diag (-\Phi,\Phi,\Phi)  \label{eqn:nongeophi}.
\end{equation}
and
\be
 L_{20} \ne 0, \qquad L_{30}=L_{40}=0.
\ee
The Weyl tensor is fully determined by the single scalar $\Phi$.
If $\Phi=0$ then the Weyl tensor vanishes, in which case the spacetime is Minkowski or (anti-) de Sitter, i.e., case (i) of the theorem. Henceforth we assume $ \Phi \ne 0$.
Since there is only one eigenvalue $-\Phi$, we know immediately (equation \eref{eqn:PhiIJ}) that the constant curvature submanifolds of theorem \ref{thm:submfd} must be 3-dimensional.

The Weyl tensor is sufficiently constrained that we can now solve completely the Bianchi equations (\ref{eqn:B8}-\ref{eqn:B10}).  The results of Section \ref{sec:submfd} still apply here, and we will make use of them below.  Indices $\al,\ba$ take values $3,4$, consistent with previous sections.

Firstly, the $\al$ component of \eref{eqn:B1}, combined with (\ref{eqn:Nalpha0},\ref{eqn:MalphaIJ}), gives
\be
  L_{\al 1} = N_{\al 0} =  \M{2}_{\al 2} = \frac{\del_\al \Phi}{4\Phi}.
\ee
Also, the $\al\ba$ components of \eref{eqn:B4} and \eref{eqn:B5}, and the $\al\ba2$ component of \eref{eqn:B7} give, after using \eref{eqn:constphi}, that
\be
  N_{\al\ba} = 0,\quad L_{\al\ba} = 0 \eqand \M{2}_{\al\ba}=0
\ee
respectively.

This now leaves us with the following results, for some unknown $L_{2a}$, $N_{2a}$, $\Phi$:
\be\label{eqn:phiresults}
  D\Phi = 0, \quad \quad \Delta \Phi = 0,\quad\quad \del_2 \Phi = 0,
\ee
\begin{equation}\label{eqn:Lresults}
 \fl L_{ij} = \left( \begin{array}{ccc}
                   L_{22} & L_{23} & L_{24}\\
                   0 & 0 & 0\\
                   0 & 0 & 0
                  \end{array}\right) , \quad
  L_{i0} = \left( \begin{array}{c}
                   L_{20}\\ 0\\ 0
                  \end{array}\right), \quad
  L_{i1} =   \left( \begin{array}{c}
                   L_{21}\\ \del_3\Phi /(4\Phi) \\ \del_4\Phi/(4\Phi)
                  \end{array} \right),
\end{equation}
\begin{equation}\label{eqn:Nresults}
 \fl N_{ij} = \left( \begin{array}{ccc}
                   N_{22} & N_{23} & N_{24}\\
                   0 & 0 & 0\\
                   0 & 0 & 0
                  \end{array}\right) , \quad
  N_{i1} = \left( \begin{array}{c}
                   N_{21}\\ 0\\ 0
                  \end{array}\right), \quad
  N_{i0} =  \left( \begin{array}{c}
                    N_{20}\\ \del_3 \Phi /(4\Phi) \\ \del_4\Phi/(4\Phi)
                  \end{array} \right),
\end{equation}
\begin{equation}\label{eqn:M2results}
 \fl \M{2}_{ij} =  \left( \begin{array}{ccc}
                   0 & 0 & 0\\
                   \del_3\Phi/(4\Phi) & 0 & 0\\
                   \del_4\Phi/(4\Phi) & 0 & 0
                  \end{array}\right) , \quad
  \M{2}_{i1} = \left( \begin{array}{c}
                   0\\ 0\\ 0
                  \end{array}\right), \quad
  \M{2}_{i0} = \left( \begin{array}{c}
                   0\\ 0\\ 0
                  \end{array}\right).
\end{equation}
Furthermore, inserting equations (\ref{eqn:phiresults}-\ref{eqn:M2results}) into (\ref{eqn:B8}-\ref{eqn:B10}), we find that this is sufficient to satisfy all of them, with no further restrictions.  

Next we shall show that the 5d spacetime must be a {\it warped product}.  We shall do this by showing that it is conformal to a product spacetime.  To this end, let $h^a{}_b$ denote the tensor that projects onto the 3d submanifolds, i.e.,
\be
 h^a{}_b = \ell^a n_b + n^a \ell_b + m_{(2)}^a m^{(2)}_b.
\ee
 Now define
\begin{equation}
  H_{abc} \equiv \nabla_c h_{ab}.
\end{equation}
Using the above results, the only non-vanishing components of this are
\begin{equation}
  H_{301}=H_{310}=H_{322}= \frac{\del_3 \Phi}{4\Phi} \eqand
  H_{401}=H_{410}=H_{422}= \frac{\del_4 \Phi}{4\Phi}
\end{equation}
as well as those related to these components by the symmetry in the first two indices. Now consider a conformally related spacetime, with metric
\be
 \tilde{g} = |\Phi|^{1/2} g.
\ee
Let $\tilde{\nabla}$ denote the Levi-Civita connection in the new spacetime. Using the relation between $\tilde{\nabla}$ and $\nabla$, and the above results for $H_{abc}$, we find that
\be
 \tilde{\nabla}_c h^a{}_b = 0.
\ee
However, this is the necessary and sufficient condition for the new spacetime to be {\it decomposable} \cite{Exact}.  That is, there exist coordinates $(x^A,y^\alpha)$ so that the metric takes the form
\be
 d\tilde{s}^2 = \tilde{g}_{AB}(x) dx^A dx^B + \tilde{g}_{\alpha\beta}(y) dy^\alpha dy^\beta,
\ee
where $A,B=0,1,2$ and $\al,\ba=3,4$ are coordinate indices only for this equation, and the remainder of this section. The 3d submanifolds are surfaces of constant $y^\alpha$. These are orthogonal to 2d submanifolds of constant $x^A$.
In this coordinate chart, equations \eref{eqn:phiresults} reduce to $\Phi=\Phi(y)$. We now see that the physical metric is a warped product:
\be
   ds^2 = |\Phi(y)|^{-1/2} \tilde{g}_{AB}(x) dx^A dx^B + g_{\alpha\beta}(y) dy^\alpha dy^\beta,
\ee
where $g_{\alpha\beta}(y) = |\Phi(y)|^{-1/2} \tilde{g}_{\alpha\beta}(y)$. The surfaces of constant $y^\alpha$ are constant curvature, so $\tilde{g}_{AB}(x)$ is the metric of 3d Minkowski or (anti-)de Sitter spacetime. We now see that the symmetries of the constant curvature submanifolds extend to symmetries of the full spacetime. Hence we can apply Birkhoff's theorem to deduce that the 5d spacetime must be isometric to either (ii) (if $\Phi$ is constant) or (iii) in the statement of the theorem. $\Box$

\subsection{Comments on the 6d case}

Consider a 6d Einstein spacetime admitting a non-geodesic multiple WAND. Let us use the same notation as we did in the proof of Proposition \ref{prop:typeD}. We already know some of the components of $C_{ijkl}$ from equation \eref{eqn:C2i2j}. The remaining components $C_{\hi\hj\hk\hl}$ have the symmetries of the Riemann tensor in 3d, hence they are completely determined by their trace $C_{\hi\hj\hk\hj}=-3\Phi_{\hi\hk}$ (see \eref{eqn:trace}). This gives
\be \label{eqn:6weyl}
 C_{\hi\hj\hk\hl} = -6 ( \del_{\hi[\hk} \Phi_{\hl]\hj} - \del_{\hj[\hk} \Phi_{\hl]\hi} ) + 6\Phi \del_{\hi [\hk} \del_{\hl]\hj}
\ee
Together with Proposition \ref{prop:typeD}, this implies that the Weyl tensor is fully determined by $\Phi_{ij}$. We can substitute this into equation \eref{eqn:B8} to learn that the constant curvature submanifolds of theorem \ref{thm:submfd} have dimension three {\it unless} (i) the eigenvalues of $\Phi_{ij}$ are $-\Phi,-\Phi,3\Phi/2,3\Phi/2$ with $\Phi \ne 0$, in which case they have dimension four; or (ii) $\Phi_{ij}=0$, in which case the spacetime is type O (i.e. Minkowski or (anti-)de Sitter spacetime).

The case of the foliation by 4d submanifolds can be analyzed using a similar method to the proof of theorem \ref{thm:list}. The result is that the spacetime must be either a direct product $dS_4 \times S^2$ or $AdS_4 \times H^2$, or a spacetime with metric 
\be
  ds^2 = r^2 d\tilde{s}_4^2 + \frac{dr^2}{U(r)} + U(r) dz^2, \qquad  U(r) = k - \frac{m}{r^3} -  \frac{\La}{5}r^2, 
\ee
where $m \ne 0$, $k \in \{1,0,-1\}$, $d\tilde{s}_4^2$ is the metric of a 4d Lorentzian space of constant curvature (i.e. 4d Minkowski or (anti-)de Sitter) with Ricci scalar $12k$, and the coordinate $r$ takes values such that $U(r)>0$. These solutions are the 6d analogues of cases (ii) and (iii) of theorem \ref{thm:list}. 

Now consider the case in which the constant curvature submanifolds are three-dimensional. In this case, we might hope to prove that the distribution orthogonal to these submanifolds is integrable.  Here, coordinates could be introduced so that the metric takes the form
\be
\label{eqn:6dnongeowand}
ds^2 =  F(x,y)^2 g_{AB}(x) dx^A dx^B + g_{\alpha\beta}(x,y) dy^\alpha dy^\beta,
\ee
where $A,B$ range from $0$ to $2$ and $\alpha,\beta$ range from $3$ to $5$ and the surfaces of constant $y^\alpha$ are the constant curvature submanifolds. The constant curvature condition implies that the coordinates $x^A$ can be chosen so that
\be
 F(x,y)^2 g_{AB}(x) = \frac{\eta_{AB}}{(a(y) \eta_{CD}x^C x^D + b_C(y) x^C + c(y))^2},
\ee
for some $a(y)$, $b_C(y)$ and $c(y)$, where $\eta_{AB}$ is the 3d Minkowski metric. Note that it is not obvious that the symmetries of the constant curvature submanifolds extend to symmetries of the spacetime.

Using the Bianchi identity, we are able to prove that the distribution orthogonal to the constant curvature submanifolds is indeed integrable {\it except} when $\Phi_{ij}$ has eigenvalues $0,0,\phi,-\phi$ for some scalar $\phi \ne 0$ (this implies $\Phi=0$). We have not made any progress in analyzing this exceptional case so we shall not give further details here. In more than six dimensions, it seems likely that the distribution orthogonal to the submanifolds of constant curvature will be non-integrable except in special cases.

\section*{Acknowledgments}
MND is supported by the Science and Technology Facilities Council. HSR is a Royal Society University Research Fellow.

\appendix

\section{Bianchi Equations for Type II spacetimes with $\Phi_{[ij]}=0$}
Here we present the Bianchi identities 
\[ \nabla_{[\tau|} C_{\mu\nu|\rho\sigma]} = 0 \]
projected into a Weyl-aligned null frame in the case of a Type II (or more special) spacetime, assuming that $\Phi_{[ij]}=0$.  This condition is always satisfied when $\lb$ is a non-geodesic multiple WAND.  The equations were obtained from \cite{Bianchi}, Appendix B.  Note that the Type D equations can be obtained by setting $\Psi_i$, $\Psi_{ijk}$ and $\Psi_{ij}$ to zero.

Here, braces $\{\}$ surrounding 3 indices denote a sum, with
\[ T_{\{abc\}} \equiv T_{abc}+T_{bca}+T_{cab} \]
and hence
\[ T_{[abc]} = \frac{1}{6}(T_{\{abc\}}-T_{\{bac\}} ) \eqand T_{(abc)} = \frac{1}{6}(T_{\{abc\}}+T_{\{bac\}} ).\]
The equations are ordered by their boost weights, as follows:\\
{\bf Boost weight +2:}
\begin{eqnarray} \label{eqn:B8}
  \fl\quad\quad 0 &=& (2\Phi_{i[j}\del_{k]s} - C_{isjk})L_{s0}
\end{eqnarray}
{\bf Boost weight +1:}
\begin{eqnarray}\label{eqn:B3}
  \fl\quad\quad 0 &=& 2(\Phi_{[i|s} + \Phi \del_{[i|s}) L_{s|j]} + 2\Psi_{[i} L_{j]0} + \Psi_{sij} L_{s0}
\end{eqnarray}
\begin{eqnarray}\label{eqn:B5}
  \fl\quad\quad D \Phi_{ij} &=& -(\Phi_{is} + \Phi \del_{is}) L_{sj} - 2\Phi_{(i|s}\M{s}_{|j)0}
                                   + (\Psi_j \del_{is} - \Psi_{jis}) L_{s0}
\end{eqnarray}
\begin{eqnarray}\label{eqn:B12} 
  \fl\quad\quad -D C_{ijkl} &=& - 2\Phi_{i[k|} L_{j|l]} + 2\Phi_{j[k|} L_{i|l]} 
                               + 2C_{ij[k|s} L_{s|l]} + 2 C_{[i|skl} \M{s}_{|j]0} + 2 C_{ij[k|s} \M{s}_{|l]0}\nonumber\\ 
                          &&       - 2 \Psi_{[i|kl} L_{|j]0} - 2 \Psi_{[k|ij} L_{|l]0}
\end{eqnarray}
\begin{eqnarray}\label{eqn:B15} 
  \fl\quad\quad 0 &=& -\Phi_{i\{j}L_{kl\}} + \Phi_{i\{j}L_{lk\}} + C_{is\{jk|}L_{s|l\}}
\end{eqnarray}
{\bf Boost weight 0:}
\begin{eqnarray}\label{eqn:B1}
  \fl\quad\quad D \Psi_i - \del_i \Phi &=& -(\Phi_{is}+\Phi\del_{is}) (L_{s1}+N_{s0})
            - \Psi_s ( \del_{si} L_{10} + 2 L_{si} + \M{s}_{i0}) + \Psi_{is} L_{s0}
\end{eqnarray}
\begin{eqnarray}\label{eqn:B6}
  \fl\quad\quad -D \Psi_{ijk} &=& (2 \Phi_{i[j} \del_{k]s} - C_{isjk}) (L_{s1}-N_{s0}) + 
                            2 \Psi_{[j} L_{k]i} + \Psi_{ijk} L_{10} + \Psi_{sjk}L_{si} \nonumber\\
                           && + \Psi_{i[j|s} \M{s}_{|k]0} + \Psi_{sjk}\M{s}_{i0} - 2\Psi_{i[j}L_{k]0}
\end{eqnarray}
\begin{eqnarray}\label{eqn:B7} 
  \fl\quad\quad -2 \del_{[j}\Phi_{k]i} &=& (2 \Phi_{i[j}\del_{k]s} - C_{isjk}) L_{s1}                                                                                   - 2\Phi_{is}\M{s}_{[jk]} - 2\Phi_{s[j|} \M{s}_{i|k]}\nonumber\\
                                  &&            + (2\Psi_{[j|} \del_{is} - 2 \Psi_{[j|is}) L_{s|k]}
\end{eqnarray}
\begin{eqnarray}\label{eqn:B9}  
\fl\quad\quad - 2 \del_{[j}\Phi_{k]i} + D \Psi_{ijk} &=& 
          (2\Phi_{i[j}\del_{k]s} - C_{isjk})N_{s0} - 2\Phi_{[j|s} \M{s}_{i|k]} - 2\Phi_{is} \M{s}_{[jk]}\nonumber\\ 
    &&    + 2 (\Psi_i \del_{[j|s} - \Psi_{i[j|s}) L_{s|k]} - \Psi_{ijk} L_{10}
          - 2\Psi_{i[j|s}\M{s}_{|k]0} \nonumber\\
      &&                   - \Psi_{sjk}\M{s}_{i0} + 2 \Psi_{i[j}L_{k]0}
\end{eqnarray}
\begin{eqnarray}\label{eqn:B11}
  \fl\quad\quad 0 &=& \Psi_{\{ i} L_{jk\}} - \Psi_{\{ i} L_{kj\}} + \Psi_{s\{ ij|} L_{s|k\}}
\end{eqnarray}
\begin{eqnarray}\label{eqn:B16}
  \fl\quad\quad -\del_{\{k|} C_{ij|lm\} } &=&  - \Psi_{i\{ kl|} L_{j|m \}} + \Psi_{j\{ kl|} L_{i|m \}}
                                                - \Psi_{\{k| ij} L_{|lm\}} + \Psi_{\{k| ij} L_{|ml\}} \nonumber \\
        &&                                        + C_{ij\{ k|s} \M{s}_{|lm\}} - C_{ij\{ k|s} \M{s}_{|ml\}} + 
                                                C_{is\{ kl|} \M{s}_{j|m\}} - C_{js\{ kl|} \M{s}_{i|m\}}
\end{eqnarray}
{\bf Boost weight -1:}
\begin{eqnarray}\label{eqn:B2}
  \fl\quad\quad - 2 \del_{[i}\Psi_{j]} &=& -2(\Phi_{[i|s}+\Phi \del_{[i|s}) N_{s|j]}\nonumber \\ 
  &&  + 2 \Psi_{[i}(L_{j]1}-L_{1|j]})+ \Psi_{sij}L_{s1} - 2 \Psi_s \M{s}_{[ij]}+ 2 \Psi_{[i|s}L_{s|j]}
\end{eqnarray}
\begin{eqnarray}\label{eqn:B4}
  \fl\quad\quad - \Del \Phi_{ij} - \del_{j}\Psi_i + D \Psi_{ij}  &=& 
                  (\Phi_{is} + \Phi \del_{is}) N_{sj} + 2\Phi_{(i|s}\M{s}_{|j)1} + \Psi_{ijs} L_{s1} \nonumber\\
          &&      + \Psi_i (L_{1j}-L_{j1})   
                  - 2\Psi_{(i} N_{j) 0} - 2\Psi_{(ij)s} N_{s0} + \Psi_s \M{s}_{ij}\nonumber\\
          &&       - \Psi_{is} L_{sj} - 2\Psi_{ij} L_{10} - 2\Psi_{(i|s} \M{s}_{|j)0}
\end{eqnarray}
\begin{eqnarray}\label{eqn:B13}  
\fl\quad\quad -\Del C_{ijkl} + 2 \del_{[k}\Psi_{l]ij} &=& 
                                2 C_{ij[k|s}N_{s|l]} + 2 C_{ij[k|s} \M{s}_{|l]1} + 2C_{[i|skl} \M{s}_{|j]1}\nonumber\\
            &&      -2\Phi_{i[k|}N_{j|l]} +  2\Phi_{j[k|}N_{i|l]}
                                -2\Psi_{[i|kl}L_{|j]1} + 2\Psi_{[k|ij} L_{|l]1} \nonumber\\
        && + 2\Psi_{[k|ij} L_{1|l]} + 2\Psi_{sij} \M{s}_{[kl]}
                                + 2\Psi_{k[i|s}\M{s}_{|j]l} - 2\Psi_{l[i|s} \M{s}_{|j]k} \nonumber\\ 
       && - 2\Psi_{i[k|} L_{j|l]} + 2\Psi_{j[k} L_{i|l]}
\end{eqnarray}
\begin{eqnarray}\label{eqn:B14}
\fl\quad\quad -\del_{\{ j|} \Psi_{ i|kl\} } &=& 
          - \Phi_{i\{j|} N_{|kl\}} + \Phi_{i\{j|} N_{|lk\}} + C_{is \{ jk|} N_{s|l\}}  + \Psi_{i\{ jk|} L_{1|l\}}\nonumber \\
          &&          + \Psi_{i\{ j|s}(\M{s}_{|kl\}} - \M{s}_{|lk\}})+ \Psi_{s\{ jk|} \M{s}_{i|l\}}
                      - \Psi_{i\{ j|} (L_{|kl\} }-L_{|lk\} } )
\end{eqnarray}
{\bf Boost weight -2:}
\begin{eqnarray}\label{eqn:B10}
  \fl\quad\quad \Del \Psi_{ijk} - 2 \del_{[j}\Psi_{k]i} &=& (2\Phi_{i[j} \del_{k]s}-C_{isjk})N_{s1}
                       -2(\Psi_{[j|} \del_{is} + \Psi_i\del_{[j|s} + \Psi_{i[j|s} + \Psi_{[j|is}) N_{s|k]} \nonumber \\
    && - \Psi_{ijk} L_{11} - 2\Psi_{i[j|s} \M{s}_{|k]1} - \Psi_{sjk} \M{s}_{i1}  \nonumber\\
    && + 2 \Psi_{i[j} L_{k]1} - 4 \Psi_{i[j|} L_{1|k]} -2\Psi_{is} \M{s}_{[jk]} - 2\Psi_{[j|s}\M{s}_{i|k]}
\end{eqnarray}

\end{document}